\documentclass[aps,prd,twocolumn,superscriptaddress]{revtex4-1}%
\usepackage{epsf,epsfig}
\usepackage[utf8]{inputenc}
\usepackage{amssymb,amsmath,amsfonts}
\usepackage{color}
\usepackage{slashed}
\usepackage{braket}
\usepackage[colorlinks=true]{hyperref}  
\hypersetup{
    bookmarks=true,         
    unicode=false,          
    pdftoolbar=true,        
    pdfmenubar=true,        
    pdffitwindow=false,     
    pdfstartview={FitH},    
    colorlinks=true,       
    linkcolor=magenta, 
    citecolor=blue,        
    filecolor=magenta,      
    urlcolor=cyan           
} 

\newcommand{\bea}{\begin{eqnarray}}
\newcommand{\eea}{\end{eqnarray}}
\newcommand{\ba}{\begin{eqnarray}}
\newcommand{\ea}{\end{eqnarray}}

\newcommand{\beq}{\begin{equation}}
\newcommand{\eeq}{\end{equation}}
\newcommand{\beqa}{\begin{eqnarray}}
\newcommand{\eeqa}{\end{eqnarray}}
\newcommand{\beqar}{\begin{eqnarray*}}
	\newcommand{\eeqar}{\end{eqnarray*}}

\newcommand{\ssc}{\scriptscriptstyle}
\newcommand{\eg}{{\it e.g.,}\ }
\newcommand{\ie}{{\it i.e.,}\ }

\newcommand{\fin}{f_\infty}

\usepackage{color}

\newcommand{\ctt}{C_{\ssc T}}
\newcommand{\ctte}{C^{\rm \ssc E}_{\ssc T}}

\newcommand{\req}[1]{(\ref{#1})} 

\begin{document}

\title{Universality of squashed-sphere partition functions}
\author{Pablo Bueno}
\affiliation{Instituut voor Theoretische Fysica, KU Leuven,
	Celestijnenlaan 200D, B-3001 Leuven, Belgium}

\author{Pablo A. Cano}
\affiliation{Instituto de F\'isica Te\'orica UAM/CSIC,
	C/ Nicol\'as Cabrera,13-15, C.U. Cantoblanco, 28049 Madrid, Spain}

\author{Robie A. Hennigar} 
\affiliation{Department of Physics and Astronomy, University of Waterloo,
	Waterloo, Ontario, Canada, N2L 3G1 }

\author{Robert B. Mann}
\affiliation{Department of Physics and Astronomy, University of Waterloo,
	Waterloo, Ontario, Canada, N2L 3G1 }


\begin{abstract}

We present several results concerning the free energy of odd-dimensional conformal field theories (CFTs) on squashed spheres. First, 
we propose a formula which computes this quantity for holographic CFTs dual to higher-curvature gravities with second-order linearized equations of motion. As opposed to standard on-shell action methods for Taub geometries, our formula only involves a simple evaluation of the corresponding bulk Lagrangian on an auxiliary pure-AdS space. The expression is closely related to the function determining the possible AdS vacua of the bulk theory in question, which we argue to
act as a generating functional from which correlation functions of the boundary stress tensor can be easily characterized.   
Finally, based on holographic results and free-field numerical calculations, we conjecture that the subleading term in the squashing-parameter free-energy expansion is universally controlled by the stress-tensor three-point function charge $t_4$ for general $(2+1)$-dimensional CFTs.


\end{abstract}
\maketitle


Euclidean conformal field theories (CFTs) coupled to background fields can be used to learn important lessons about the dynamics of the theory in question.  A prototypical example corresponds to supersymmetric CFTs, where localization techniques have allowed for notable progress --- see \eg \cite{Pestun:2016zxk}. For non-supersymmetric theories, a natural possibility consists in coupling the theory to curved background metrics. This approach has produced some exact and universal results valid for general CFTs \cite{Bobev:2017asb,Fischetti:2017sut} and has found various applications, \eg in holographic cosmology \cite{Anninos:2012ft,Conti:2017pqc,Hertog:2017ymy,Hawking:2017wrd}. Particularly interesting is the case of spherical backgrounds, whose partition functions --- equivalently, free energies: $F_{\mathbb{S}^d}=- \log |Z_{\mathbb{S}^d}|$ --- 
have been conjectured to be renormalization-group monotones for general odd-dimensional QFTs \cite{Klebanov:2011gs,Casini:2012ei,Pufu:2016zxm}.  

In this letter, we will consider CFTs on deformed spheres and study the effect that such deformations have on $F$. The focus will be on a particular class of squashed spheres, $\mathbb{S}_{\varepsilon}^d$, which preserve a large subgroup of isometries of the round ones \footnote{In particular, \req{squa} preserves a SU$(\frac{d+1}{2})\times$U$(1)$ subgroup of the usual SO$(d+1)$ preserved by the usual round-sphere metric in $d$-dimensions.}. In particular, they are characterized by being Hopf fibrations over the complex projective space $\mathbb{CP}^{k}$ ($k\equiv (d-1)/2$), namely, $\mathbb{S}^1\hookrightarrow \mathbb{S}_{\varepsilon}^d\rightarrow \mathbb{CP}^k$. The metric on these  squashed-spheres is given by
\begin{equation}\label{squa}
ds^2_{\mathbb{S}_{\varepsilon}^d}=\frac{ds^2_{\mathbb{CP}^k}}{(d+1)}+(1+\varepsilon)\left(d\psi+\frac{A_{\mathbb{CP}^k}}{(d+1)}\right)^2\, ,
\end{equation}
where $\psi$ is a periodic coordinate which parametrizes the $\mathbb{S}^1$, $ds^2_{\mathbb{CP}^k}$ is the Einstein metric on $\mathbb{CP}^k$ normalized so that $R_{ij}=g_{ij}$, and $J=dA_{\mathbb{CP}^k}$ is the K\"ahler form on $\mathbb{CP}^k$. The parameter $\varepsilon$ measures the degree of squashing of the sphere and, in principle, it can take values in the domain $\varepsilon \in (-1,+\infty)$, the round-sphere corresponding to $\varepsilon=0$. In $d=3$, which is the simplest case, $\mathbb{CP}^1 \cong \mathbb{S}^2$, and we can write $ds^2_{\mathbb{S}^2}=d\theta^2+\sin^2\theta d\phi^2$, $A_{\mathbb{S}^2}=2\cos \theta d\phi$ in standard spherical coordinates.

This class of squashed spheres can be easily studied holographically \cite{Hawking:1998ct,Dowker:1998pi,Chamblin:1998pz,Emparan:1999pm,Hartnoll:2005yc,Bobev:2016sap}, as the relevant bulk geometries belong to the well-known AdS-Taub-NUT/bolt family. Our first main result --- see \req{fee0e} --- is a universal formula for the free-energy of a broad class of holographic CFTs on squashed-spheres. The formula is automatically UV-finite and, in fact, does not require knowing the corresponding NUT solutions explicitly. It holds for higher-curvature bulk theories with second-order linearized equations of motion, correctly reproducing all known results available for such theories, and passes several consistency checks coming from field theory considerations.
 Our second result --- see \req{3conj} --- is an expression for the subleading term in the small squashing-parameter expansion of $F_{\mathbb{S}_{\varepsilon}^{d}}$ which, based on holographic and free field calculations we conjecture to be controlled by the stress-tensor three-point function coefficient $t_4$ for general CFTs. 
As an additional consequence of our results in the holographic context, we observe that, for the class of bulk theories just described, the function that determines the possible AdS vacua of the theory --- see \req{hgo} --- acts as a generating functional for the boundary stress-tensor, in the sense that we can easily characterize its correlators by taking trivial derivatives of such function, drastically simplifying the standard holographic calculations  --- see \req{cte}, \req{tte}, \req{0p} and \req{conne}.


\textbf{Higher-order gravities and holography on squashed-spheres:}
AdS/CFT \cite{Maldacena,Witten,Gubser} provides a powerful playground for exploring the physics of strongly coupled CFTs. In some cases, the possibility of mapping intractable field-theoretical calculations into manageable ones involving gravity techniques allows for the identification of universal properties valid for completely general CFTs.  In this context, higher-curvature gravities turn out to be very useful, as they define holographic toy models for which many explicit calculations 
can be performed explicitly. The idea is that, if a certain property is valid for general theories, it should also hold for these models. This approach has been successfully used before, \eg in the identification of monotonicity theorems in various dimensions \cite{Myers:2010xs,Myers:2010tj}, or in the characterization of entanglement entropy universal terms \cite{Bueno1,Bueno2,Mezei:2014zla,Chu:2016tps}. Naturally, particular higher-curvature interactions generically appear as stringy corrections to the effective actions of top-down models admitting holographic duals \cite{Gross:1986mw}. For the purposes just described, however, it is more useful to consider bulk models which are particularly amenable to holographic calculations --- see \eg \cite{Camanho:2009hu,Buchel:2009sk,Myers:2010jv,deBoer:2009gx,Camanho:2013pda,HoloECG}. 

The Lagrangian of such kind of models can be generally written, in $(d+1)$ bulk dimensions, as 
\begin{equation}\label{hog}
\mathcal{L}=\frac{1}{16\pi G}\left[\frac{d(d-1)}{L^2}+R+\sum_{n=2} \mu_n L^{2(n-1)}\mathfrak{R}_{(n)}\right]\, ,
\end{equation}
where $L$ is some length scale,  $G$ is Newton's constant, the $\mu_{n}$ are dimensionless couplings, and the $\mathfrak{R}_{(n)}$ stand for the higher-curvature terms, constructed from  linear combinations of order-$n$ curvature invariants. 
The AdS vacua of any theory of the form \req{hog} can be obtained by solving \cite{Aspects}
\begin{equation}\label{hinf}
h(f_{\infty})\equiv\frac{16\pi G L^2}{d(d-1)}\left[\mathcal{L}(f_{\infty})-\frac{2f_{\infty}}{(d+1)}\mathcal{L}'(f_{\infty})\right]=0\, ,
\end{equation}
where $\mathcal{L}(f_{\infty})$ is the on-shell Lagrangian on pure AdS$_{(d+1)}$ with radius $L/\sqrt{f_{\infty}}$. This 
 can be easily obtained evaluating all Riemann tensors in \req{hog} as $R_{abcd}=-f_{\infty}/L^2 (g_{ac}g_{bd}-g_{ad}g_{bc})$. Also, $\mathcal{L}'(f_{\infty})\equiv d\mathcal{L}(f_{\infty})/df_{\infty}$.  It is easy to argue that the $\mathfrak{R}_{(n)}$ can always be normalized so that the function $h(f_{\infty})$ in \req{hinf} reduces to the form\footnote{The special case $n=(d+1)/2$ is automatically excluded from the sum, as no invariant of that order contributes to the vacua equation.} 
\begin{equation}\label{hgo}
h(f_{\infty})=1-f_{\infty}+\sum_{n=2} \mu_{n} f_{\infty}^n\, .
\end{equation}
For Einstein gravity one just finds $f_{\infty}=1$, and the action scale $L$ coincides with the AdS radius. 

We further restrict \req{hog} to the particular subclass of theories whose linearized equations on maximally symmetric backgrounds are second-order 
\footnote{Namely, we restrict to those for which the linearized equations take the form $G_{ab}^{\rm L}=8\pi G_{\rm eff} T_{ab}$, where $G_{ab}^{\rm L}$ is the linearized Einstein tensor, $T_{ab}$ is  some possible matter stress-tensor, and $G_{\rm eff}$ is the effective Newton constant}. This subclass --- which we shall refer to as \emph{Einstein-like}  \cite{Aspects} --- contains infinitely many theories and includes: all Lovelock  \cite{Lovelock1,Lovelock2} and some $f($Lovelock$)$ theories \cite{Love}, Quasi-topological gravity \cite{Quasi2,Quasi} and its higher-curvature extensions \cite{Dehghani:2011vu,Cisterna:2017umf}, Einsteinian cubic gravity in general dimensions \cite{PabloPablo}, and Generalized Quasi-topological gravity \cite{Hennigar:2017ego}, among others \cite{Karasu:2016ifk,Li:2017ncu,Li:2017txk}. The vast majority of all known theories of the form \req{hog} admitting non-trivial black hole and Taub solutions belong to this class. 


As we show here, the function $h(f_{\infty})$ contains a surprisingly great deal of additional nontrivial information for Einstein-like theories.
First, given one such theory, it determines the effective gravitational constant through $G_{\rm  eff}=-G/h'(f_{\infty})$ --- see the Supplement for a proof. From the dual CFT point of view, this translates into the following relation with the charge $\ctt$, which fully characterizes the CFT  stress-tensor two-point function\footnote{Conformal invariance completely constrains the correlator $ \braket{ T_{\mu\nu}(x)\, T_{\lambda\rho}(0)} $ up to a theory-dependent quantity, customarily denoted $\ctt$, as
$ \braket{ T_{\mu\nu}(x)\, T_{\lambda\rho}(0) } =\ctt\, \mathcal I_{\mu\nu,\lambda\rho}(x) / |x|^{2d}\,, $ where $\mathcal I_{\mu\nu,\lambda\rho}$ is a fixed dimensionless tensor structure \cite{Osborn:1993cr}.}\footnote{Eq. \req{cte} was previously proven in the particular case of Lovelock theories in \cite{Camanho:2010ru,Camanho:2013pda}.} 
\begin{equation}\label{cte}
\ctt = -h'(f_{\infty})\ctte\, , 
\end{equation}
where $\ctte$ stands for the Einstein gravity result\footnote{Observe that our convention for $\ctt$ differs from that in \cite{Bobev:2017asb} by a factor $1/\mathcal{S}_d^2=\Gamma[d/2]^2/(4\pi^2)$. It agrees, however, with the convention in \cite{Myers:2010tj,Bueno2,Buchel:2009sk,Myers:2010jv}. Note also that it is customary to write Einstein gravity results in terms of $L/\sqrt{f_{\infty}}$, instead of $L$ alone. This is irrelevant for Einstein gravity itself, for which $f_{\infty}=1$, but needs to be kept in mind for higher-order theories.} 
\begin{equation}\label{cte2}
\ctte=\frac{\Gamma[d+2] (L/\sqrt{f_{\infty}})^{d-1}}{8\pi^{\frac{d+2}{2}}(d-1)\Gamma\left[\frac{d}{2} \right]G}\, .
\end{equation}



In AdS/CFT, the semiclassical partition function is exponentially dominated by the bulk geometry with the smallest on-shell action satisfying the appropriate boundary conditions. Hence, the free energy of the CFT can be accessed from the regularized on-shell action of the  bulk theory evaluated on the corresponding gravity solution \cite{Aharony:1999ti}.
  When the boundary geometry is a squashed-sphere of the form \req{squa}, the relevant bulk solutions are of the so-called Euclidean Taub-NUT/bolt class \cite{Hawking:1998ct,Chamblin:1998pz,Emparan:1999pm}. Such solutions are characterized by the NUT charge $n$ which, on general grounds, holography maps to the squashing parameter of the boundary geometry $\varepsilon$ through
\begin{equation}\label{n-epsilon}
\frac{n^2}{L^2} =\frac{(1+\varepsilon)}{(d+1) f_{\infty} }\, .
\end{equation}

Naturally, constructing Taub solutions is a more challenging task than classifying the vacua of the theory and, in fact, only a few examples of such solutions have been constructed for Einstein-like Lagrangians of the form \req{hog}. The simplest instances in $d=3$ correspond to Einsteinian cubic gravity \cite{NewTaub2}, whose Lagrangian is given by \cite{PabloPablo}
\begin{equation}\label{ecgg}
\mathcal{L}^{\rm ECG}=\frac{1}{16\pi G}\left[\frac{6}{L^2}+R-\frac{\mu L^4}{8} \mathcal{P} \right]\, ,
\end{equation}
where $\mathcal{P}=12 R_{a\ b}^{\ c \ d}R_{c\ d}^{\ e \ f}R_{e\ f}^{\ a \ b}+R_{ab}^{cd}R_{cd}^{ef}R_{ef}^{ab}-12R_{abcd}R^{ac}R^{bd}+8R_{a}^{b}R_{b}^{c}R_{c}^{a}$ is a new cubic invariant and  $\mu$ is a dimensionless coupling. In $d\geq 5$, analytic Taub solutions have been constructed for Einstein \cite{Awad:2000gg} and Einstein-Gauss-Bonnet gravity \cite{Dehghani:2005zm,Dehghani:2006aa,Hendi:2008wq} and there have been a number of holographic applications of these solutions \cite{Astefanesei:2004kn,Clarkson:2004yp,Lee:2008yqa,Shaghoulian:2016gol}.
 Very recently, additional solutions have been discovered for other Einstein-like theories (both in $d=3$ and $d=5$) in \cite{NewTaub2}. 

 In all these cases, the thermodynamic properties of the solutions can be accessed analytically. In particular, the computation of  regularized on-shell actions can be performed after the introduction of various boundary terms and counterterms which account for the various UV divergences \cite{Chamblin:1998pz,Balasubramanian:1999re, Brihaye:2008xu, Teitelboim:1987zz,Dehghani:2011hm,HoloECG}.  As long as the solution is the dominant saddle, 
 the resulting on-shell action computes the free energy of the dual theory on a squashed sphere $\mathbb{S}^d_{\varepsilon}$. For sufficiently small $\varepsilon$, the relevant saddle is generically of the NUT type. 
 
 \textbf{A universal formula for holographic squashed-spheres free energy:}
  Rather strikingly, we observe that the following simple pattern holds in all cases: the finite (and only physically meaningful) contribution to the free energy of a holographic CFT dual to an Einstein-like higher-order gravity theory on a squashed $\mathbb{S}^d_{\varepsilon}$  can be obtained by evaluating the on-shell Lagrangian of the corresponding theory on pure AdS$_{(d+1)}$. The dependence on the squashing parameter appears encoded in the AdS radius of this auxiliary geometry, which is given by $L \sqrt{(1+\varepsilon)/f_{\infty}}$. Explicitly, we claim that for any theory of this kind the following formula holds 
\begin{equation}\label{fee0e}
F_{\mathbb{S}_{\varepsilon}^{d}}=(-1)^{\frac{(d-1)}{2}}\frac{\pi^{\frac{(d+2)}{2}} }{\Gamma\left[\frac{d+2}{2}\right]}\frac{\mathcal{L}\left[f_{\infty}/(1+\varepsilon)\right] L^{d+1}}{[f_{\infty}/(1+\varepsilon)]^{\frac{(d+1)}{2}}}\, .
\end{equation}
This expression is drastically simpler than the standard on-shell action approach, which  relies on various theory-dependent (boundary and counter-) terms to extract this finite piece. Instead, in \req{fee0e} the regularization is automatically implemented, and allows us to perform a general theory-independent analysis of the free energy of holographic CFTs on squashed-spheres.


First, note that if we set $\varepsilon=0$, we recover the result for the free energy of the theory on a round $\mathbb{S}^{d}$, which plays a crucial role in establishing monotonicity theorems, particularly in three-dimensions \cite{Klebanov:2011gs,Casini:2012ei,Pufu:2016zxm}. 
Indeed, this quantity has been argued to satisfy  $F_{\mathbb{S}^{d}} \propto \mathcal{L}(f_{\infty})$ for general higher-curvature bulk theories, 
with the proportionality coefficient precisely agreeing with the one predicted by \req{fee0e} --- see \eg \cite{Myers:2010tj,HoloECG}. 
Hence, from the boundary CFT point of view, \req{fee0e} tells us that the problem of computing $F_{\mathbb{S}_{\varepsilon}^{d}}$ for a given  Einstein-like theory, can actually be mapped to the one of evaluating the round $\mathbb{S}^d$ free energy for a different theory characterized by the same bulk Lagrangian, but different couplings $\tilde{\mu}_n$ such that $h(\tilde{f}_{\infty})=0$ is satisfied for $\tilde{f}_{\infty}\equiv f_{\infty}/(1+\varepsilon)$. 

An apparently similar connection between $\mathfrak{F}_{\mathbb{S}_{\varepsilon}^{d}}$ and  $\mathfrak{F}_{\mathbb{S}^{d}}$ was found for $d=3$, $\mathcal{N}=2$ supersymmetric CFTs in  \cite{Hama:2011ea}. However, supersymmetry requires additional background fields to be turned on besides the metric, which makes the corresponding free energies  $\mathfrak{F}_{\mathbb{S}_{\varepsilon}^{d}}$ inequivalent from our $F_{\mathbb{S}_{\varepsilon}^{d}}$ \cite{Bobev:2017asb}. Besides, the independence on the squashing parameter is generally true in the supersymmetric case, but not for $F_{\mathbb{S}_{\varepsilon}^{d}}$.


Back to the implications of \req{fee0e}, we know that the round sphere is a local extremum for the function $F_{\mathbb{S}_{\varepsilon}^{d}}$ \cite{Bobev:2017asb}, namely, $dF_{\mathbb{S}_{\varepsilon}^{d}}/d\varepsilon|_{\varepsilon=0}\equiv F_{\mathbb{S}_{\varepsilon}^{d}}'(0)=0$ for general theories. This is also nicely implemented in \req{fee0e}. Indeed, comparing with \req{hinf}, it is straightforward to show that, according to \req{fee0e}, 
$
F_{\mathbb{S}_{\varepsilon}^{d}}'(0)\propto h(f_{\infty})\, ,
$
which of course vanishes by definition, as $h(f_{\infty})=0$ is nothing but the embedding condition of AdS$_{(d+1)}$ on the corresponding theory. 
Furthermore, we know that $F_{\mathbb{S}_{\varepsilon}^{d}}''(0)$ is fully determined by the stress tensor two-point function charge $\ctt$ for general odd-dimensional CFTs \cite{Bobev:2017asb}. In particular, for $d=3$ and $d=5$, it was found (in our conventions) that\footnote{Related expressions had been previously found in the context of $d=3$, $\mathcal{N}=2$ supersymmetric CFTs  \cite{Closset:2012ru} --- see also \cite{Closset:2012vp,Closset:2012vg,Hama:2011ea,Imamura:2011wg,Martelli:2011fu}. A detailed discussion of the connection can be found in section 5.1 of \cite{Bobev:2017asb}.}
\begin{equation}\label{sw2}
F_{S^3_{\varepsilon}}''(0)=-\frac{\pi^4}{3}\ctt\, , \quad F_{S^5_{\varepsilon}}''(0)=+\frac{\pi^6}{15}\ctt\, .
\end{equation}
Now, using \req{hinf}, \req{cte} and \req{fee0e}
we find, after some manipulations,
\begin{equation}\label{genD2}
F_{\mathbb{S}_{\varepsilon}^{d}}''(0)=\frac{(-1)^{\frac{(d-1)}{2}}\pi^{d+1} (d-1)^2 }{  2\,  d!}\ctt\, .
\end{equation}

This expression reduces to the general results in \req{sw2}, which is another highly non-trivial check of \req{fee0e}. Interestingly, it provides a generalization of the universal connection between $F_{S^d_{\varepsilon}}''(0)$ and $\ctt$ which must hold for general odd-dimensional CFTs (holographic or not).

\textbf{Universal expansion on the squashing parameter:}
\begin{figure}[t]
	\includegraphics[scale=0.7]{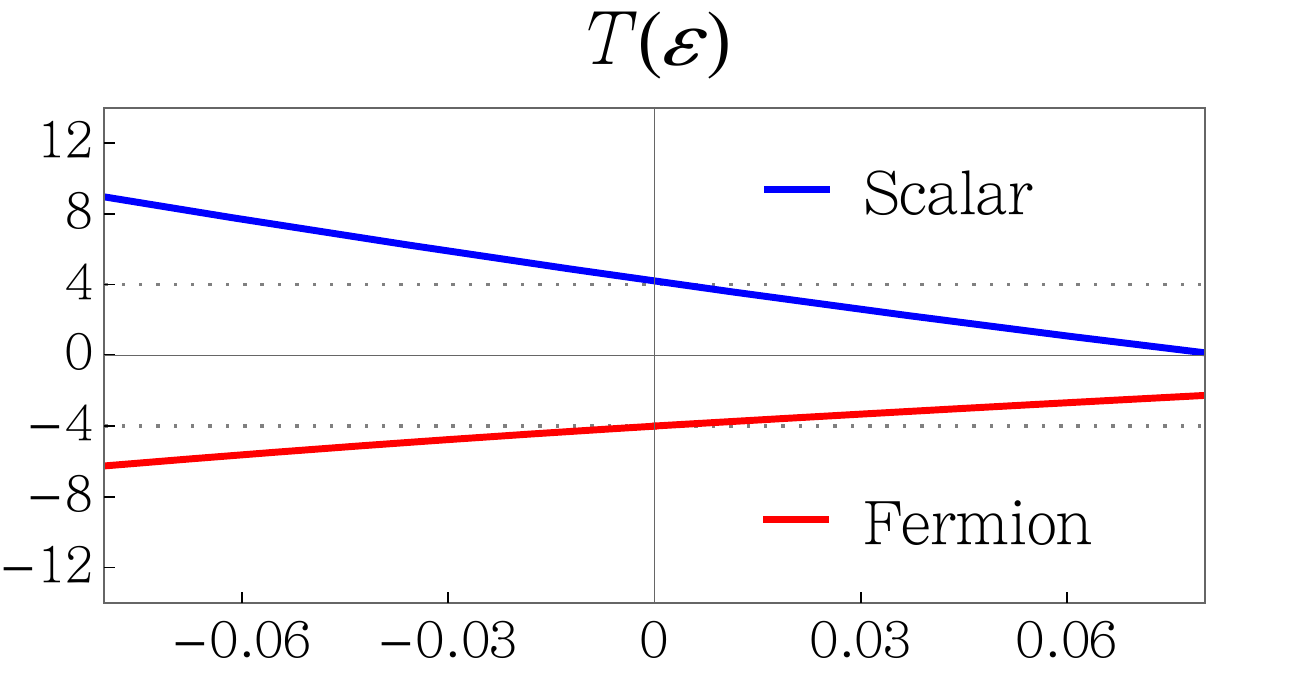}
	\caption{We plot the function $T(\varepsilon)$ defined in \req{tep} near $\varepsilon=0$ for a free scalar (blue) and a free fermion (red) using the numerical results for $F^{\rm s,f}_{\mathbb{S}_{\varepsilon}^{3}}$ obtained in \cite{Bobev:2017asb}. $T(\varepsilon=0)=t_4$ is satisfied in both cases with high accuracy, which provides strong evidence in favor of the conjectural general expression \req{3conj}. 
	}
	\label{fig.2}
\end{figure}
As we have seen, the leading term in the $\varepsilon\rightarrow 0$ expansion of $F_{\mathbb{S}_{\varepsilon}^{d}}$ is quadratic in the deformation, and proportional to the stress-tensor two-point function charge $\ctt$ for general CFTs. A question left open in \cite{Bobev:2017asb} was the possibility that the subleading term, cubic in $\varepsilon$, could present an analogous universal behavior, in the sense of being fully characterized by the corresponding three-point function charges. Since $\varepsilon$ encodes a metric deformation, one expects $F^{(n)}_{\mathbb{S}_{\varepsilon}^{d}}$ to involve integrated $n$-point functions of the stress tensor.
For general parity-even three-dimensional CFTs, the corresponding three-point function is completely fixed by conformal symmetry up to two theory-dependent quantities \cite{Osborn:1993cr}. These can be chosen to be $\ctt$, plus an additional dimensionless quantity, customarily denoted $t_4$ \cite{Hofman:2008ar}. Hence, we expect a linear combination of $\ctt$ and $\ctt t_4$ to appear in the $\mathcal{O}(\varepsilon^3)$ term.
The analysis in \cite{Bobev:2017asb} shows however that, besides these contributions, an additional correlator of the form $\braket{\frac{\delta T}{\sqrt{g}\delta g} T}$ --- which depends on additional details of the specific CFT --- appears at that order for general metric perturbations. The possibility that this term does not really contribute for certain metric perturbations, including our class of squashings, was left open.

The available partial results --- numerical for a free scalar and a free fermion, and analytic for holographic Einstein gravity --- did not suffice to provide a conclusive answer. In particular, the exact result for the free energy in holographic Einstein gravity is a polynomial of order $2$ in $\varepsilon$, namely, $F^{\rm E}_{\mathbb{S}_{\varepsilon}^{3}}=\pi L^2 (1-\varepsilon^2)/(2 G)$, which means that its Taylor expansion around $\varepsilon=0$ is trivial, and precisely ends with the quadratic piece --- which is of course controlled by $\ctt$ in agreement with \req{sw2}, as can be readily verified using \req{cte2}. 

Happily, the new Taub-NUT solutions constructed in \cite{NewTaub2} for Einsteinian cubic gravity provide us with an additional family of holographic models for which we can access the cubic contribution, and explore its possible universality by testing it against the free-field numerics.

Using the result obtained in \cite{HoloECG} for $t_4$ in holographic Einsteinian cubic gravity, we can express the squashed-sphere free energy of the corresponding dual theory 
for small values of $\varepsilon$ as 
\begin{equation}\label{3conj}
F_{\mathbb{S}_{\varepsilon}^{3}}=F_{\mathbb{S}_0^{3}}-\frac{\pi^4\ctt}{6}\varepsilon^2\left[1-\frac{t_4}{630}\varepsilon + \mathcal{O}(\varepsilon^2) \right]\, ,
\end{equation} 
where the holographic mapping between boundary  and bulk quantities is given by: $F^{\rm ECG}_{\mathbb{S}_0^{3}}=(1+3\mu f_{\infty}^2)\pi L^2/(2f_{\infty}G)$, $\ctt^{\rm ECG}=(1-3\mu f_{\infty}^2)3L^2/(\pi^3 f_{\infty}G )$ and $\ctt^{\rm ECG} t_4^{\rm ECG}=-3780\mu f_{\infty}L^2/(\pi^3 G)$, which naturally reduce to the Einstein gravity results in the $\mu \rightarrow 0$ limit. 

The leading correction to the round-sphere result agrees with the general result \req{sw2}, as it should. But now we have a nontrivial subleading piece, cubic in $\varepsilon$, and proportional to $\ctt t_4$. 
As we mentioned above, it is not obvious
that this term should not depend on additional theory-dependent quantities on general grounds. Luckily, we can use the numerical free-field results in \cite{Bobev:2017asb} to perform two highly nontrivial tests of the possible validity of \req{3conj} beyond holography. In order to do so, we study the function
 \begin{equation}\label{tep}
T(\varepsilon)\equiv \frac{630}{\varepsilon}\left[ 1+\frac{6(F_{\mathbb{S}_{\varepsilon}^{3}}-F_{\mathbb{S}_0^{3}}) }{\pi^4 \ctt \varepsilon^2} \right]
 \end{equation}
 for the conformally-coupled scalar (s) and the free Dirac fermion (f) free energies near $\varepsilon=0$. Naturally, if \req{3conj} held for these theories, we should obtain $T(\varepsilon=0)=t_4$ which, for the scalar and the fermion are respectively given by $t^{\rm s}_4=+4$ and $t^{\rm f}_4=-4$ \cite{Osborn:1993cr,Buchel:2009sk}. The result of this analysis is shown in Fig. \ref{fig.2}, where it is manifest that this is precisely satisfied in both cases --- details on the numerical methods utilized in the computation of $F^{\rm s}_{\mathbb{S}_{\varepsilon}^{3}}$, $F^{\rm f}_{\mathbb{S}_{\varepsilon}^{3}}$, and $T(\varepsilon)$ can be found in the Supplement. The extremely different nature of the theories and techniques used in deriving the holographic and free-field results make us think that this property extends to arbitrary CFTs.

\begin{itemize}
\item \textit{Conjecture:} for general three-dimensional CFTs, the subleading term in the  squashing-parameter $\varepsilon$ expansion of the free energy $F_{\mathbb{S}_{\varepsilon}^{3}}$ is universally controlled by the coefficient $t_4$ in the three-point function of the stress tensor. In particular, we conjecture that \req{3conj}
holds for general theories. 
\end{itemize}
The level of evidence provided here in favor of \req{3conj} --- involving free-field and holographic higher-order gravity calculations ---  is very similar to the one initially presented in \cite{Bueno1,Bueno2} concerning the universal relation between the entanglement entropy of almost-smooth corner regions and the charge $\ctt$, which was eventually proven for general CFTs in \cite{Faulkner:2015csl}\footnote{In contrast to \req{3conj}, however, the subleading term in the smooth-limit expansion of the corner entanglement entropy (quartic in the deformation), was later shown not to be generically controlled by the stress tensor three-point function charges in \cite{Bueno:2015ofa}.}.


If our conjecture is true, an analogous expression should hold for higher odd-dimensional squashed spheres. In that case, one would expect the $\mathcal{O}(\varepsilon^3)$ term to be controlled by some combination of $\ctt$, $t_4$ and the additional stress-tensor three-point function charge, $t_2$, which is nonvanishing for $d>3$. 




\textbf{Final comments:} 
In spite of the compelling evidence presented here in favor of our new conjectural relation \req{3conj},  performing additional checks would be very desirable. In particular, this could be further tested using the holographic duals of the set of higher-order theories constructed in \cite{PabloPablo4}. A more ambitious (and challenging) endeavour would be to prove it for general CFTs using field-theoretical techniques along the lines of \cite{Bobev:2017asb}. For this, one would need to explain why the $\braket{\frac{\delta T}{\sqrt{g}\delta g} T}$ correlator mentioned above makes no contribution in this case. 

The applicability of \req{fee0e} is of course more restrictive, as it holds only for a specific class of holographic theories. Proving it in general for such theories would also be interesting though.

Put together, \req{fee0e} and \req{3conj} would have additional consequences in the holographic context. As one can easily check, $F^{(n)}_{S^d_{\varepsilon}}(0)$ 
can always be written in terms of $(n-1)$-th (and lower) derivatives of $h(f_{\infty})$.
For example, one finds 
\begin{align}\notag
F_{\mathbb{S}^d_{\varepsilon}}^{(3)}(0)=& \, \frac{(-1)^{\frac{(d+1)}{2}}\pi^{\frac{d}{2}}(d^2-1)L^{d-1}}{16\Gamma[\frac{d}{2}]f_{\infty}^{\frac{d-1}{2}} G} \\ \label{sw3} & \cdot \left[(d-3)h'(f_{\infty})   -f_{\infty}h''(f_{\infty}) \right].
\end{align}
Then, if \req{3conj} holds for general theories, it follows that for any $d=3$ holographic higher-order gravity of the Einstein-like class\footnote{It is immediate to check that this expression yields the right $t_4$ for Einsteinian cubic gravity \req{ecgg}, for which $h(f_{\infty})=1-f_{\infty}+\mu f_{\infty}^3$.}, 
\begin{equation}\label{t4d}
t_4=210 f_{\infty} \frac{h''(f_{\infty})}{h'(f_{\infty})}\, .
\end{equation}
Hence, one would be able to obtain the coefficient $t_4$ by taking a couple of derivatives of $h(f_\infty)$. This represents a dramatic simplification with respect to the standard holographic calculations involving energy fluxes --- see \eg \cite{Hofman:2008ar,Buchel:2009sk,Myers:2010jv}. It is natural to expect that this formula generalizes to higher-dimensions. In that case, we expect an expression of the form
\begin{equation}
\label{tte}
a_{(d)} t_2 + b_{(d)} t_4 =f_{\infty} \frac{h''(f_{\infty})}{h'(f_{\infty})}\, ,
\end{equation}
to hold for general Einstein-like theories in arbitrary dimensions, for some dimension-dependent constants $a_{(d)}$ and $b_{(d)}$. Using the available results for $t_2$ and $t_4$ in $d=4$ Quasi-topological gravity \cite{Myers:2010jv} and $d\geq 4$ Gauss-Bonnet \cite{Buchel:2009sk}, it is straightforward to set: $b_{(4)}=-1/21$ and $a_{(d)}=-(d-2)(d-3)/[2d(d-1)]$. In fact, a formula equivalent to \req{tte} valid in the particular case of Lovelock theories --- for which $t_4=0$ ---  was shown to be true in \cite{Camanho:2010ru,Camanho:2013pda} for the same value of $a_{(d)}$. This provides additional support for the validity of \req{tte} for general Einstein-like theories.
It would be interesting to test the validity for such additional theories in various dimensions and, if correct in general, to determine the value of $b_{(d\geq 5)}$.

$h^{(n)}(f_\infty)$ appears to be related to the $(n+1)$-point function of the boundary stress tensor, therefore acting as some sort of generating functional. Interestingly,  the 
``zero-point function'' corresponding to the regularized round-sphere free energy $F_{\mathbb{S}^d}$ also satisfies this pattern, as it can be extracted from an integral involving $h(f_{\infty})$, namely\footnote{In even-dimensional CFTs, this expression yields --- up to a  $2(-1)^{-1/2}/\pi$ factor --- the coefficient of the universal logarithmic contribution to the corresponding round-sphere free energy, given by $(-1)^{\frac{(d-2)}{2}}4a^*$, where $a^*$ is proportional to one of the trace-anomaly charges ($a$ in $d=4$), \eg \cite{Myers:2010tj}.  }
\begin{equation}\label{0p}
F_{\mathbb{S}^d}=\frac{(-1)^{\frac{(d+1)}{2}} \pi^{\frac{d}{2}} (d+1)(d-1)L^{d-1}}{16 \Gamma\left[\frac{d}{2}\right]  G }\int^{f_{\infty}} \frac{ h(x)}{  x^{\frac{(d+3)}{2}}} dx\, .
\end{equation}
Integrating by parts in this expression, and using \req{cte} and \req{cte2}, it is possible to find the suggestive relation\footnote{Let us note that there is an ambiguity in this expression because there are many ways to express $F_{\mathbb{S}^d}$ as a function of $f_{\infty}$. There is a canonical form of $F_{\mathbb{S}^d}$ in which \req{conne} works, and this form is $F_{\mathbb{S}^d}=F^{\rm E}_{\mathbb{S}^d}\left(1+\text{polynomial in }f_{\infty}\right)$, where $F^{\rm E}_{\mathbb{S}^d}$ is the Einstein gravity result with the rescaled AdS scale $L/\sqrt{f_{\infty}}$. }
\begin{equation}\label{conne}
\ctt =\frac{(-1)^{\frac{(d-1)}{2}}  \Gamma[d+2] }{\pi^{d+1}(d-1)^2} \, f_{\infty}\left[ \frac{\partial F_{\mathbb{S}^d}}{\partial f_{\infty}} \right]\, ,
\end{equation}
which is equivalent to the one recently found in \cite{Li:2018drw}, and which, for this class of theories, connects two seemingly unrelated quantities, such as $\ctt$ and  $F_{\mathbb{S}^d}$ \footnote{In terms of $a^*$, the relation reads $\ctt= -2\Gamma[d+2]/[\pi^d (d-1)^2] \cdot f_{\infty} \left[\frac{\partial a^*}{\partial f_{\infty}}\right]$, which is valid in general (even and odd) dimensions.}.


\vspace{0.5cm}
{\it Acknowledgments.}   We would like to thank Nikolay Bobev, Alejandro Ruip\'erez and Yannick Vreys for useful discussions and comments.
The work of PB was supported by a postdoctoral fellowship from the National Science Foundation of Belgium
(FWO). The work of PAC is funded by Fundaci\'on la Caixa through a ``la Caixa - Severo Ochoa''
International pre-doctoral grant  and partially by the MINECO/FEDER, UE grant FPA2015-66793-P,
and the ``Centro de Excelencia Severo Ochoa'' Program grant SEV-2016-0597. 
RBM and RAH were supported in part by the Natural Sciences and Engineering Research Council of Canada.

\onecolumngrid  \vspace{1.3cm} 
\begin{center}  
{\Large\bf Supplementary Information}  \vspace{-0.2cm} 
\end{center} 
\appendix 
\tableofcontents

\section{Effective Newton constant in Einstein-like theories}
\label{cth}
The spectrum of general $\mathcal{L}(g^{ab},R_{acbd})$ theories on maximally symmetric backgrounds (m.s.b.) was characterized in \cite{Aspects}. For any given higher-order theory of this kind, the linearized field equations around a m.s.b. $\bar g_{ab}$ with curvature scale $\Lambda$, are given by
\begin{align}\notag
\frac{1}{2}\mathcal{E}_{ab}^{L}=&+\left[e-2\Lambda(a(D-1)+c)+(2a+c)\bar \Box\right]G_{ab}^{ L}+\left[a+2b+c\right]\left[\bar g_{ab}\bar\Box-\bar\nabla_{a}\bar\nabla_{b}\right]R^{ L}\\  &-\Lambda\left[a(D-3)-2b(D-1)-c \right] \bar g_{ab}R^{ L}
\, , \label{lineareqs}
\end{align}
where $G_{ab}^L$ and $R^L$ stand for the linearized Einstein tensor and the Ricci scalar, respectively (and $d=D-1$).  As we can see, there is a fixed theory-independent tensorial structure which is weighted by linear combinations of four theory-dependent parameters, which were denoted $a,b,c,e$ in \cite{Aspects}. These parameters determine the physical quantities of the theory --- namely, the masses of the ghost-like graviton ($m_g$) and the scalar mode ($m_s$), and the effective Newton constant $(G_{\rm eff})$ --- and  can be straightforwardly computed for a given theory following the procedure presented in \cite{Aspects}.   The relations between $a,b,c,e$ and $m_g^2,m_s^2,G_{\rm eff}$ read 
\begin{equation}\label{kafka}
m_s^2=\frac{e(D-2)-4\Lambda(a+bD(D-1)+c(D-1))}{2a+Dc+4b(D-1)}\, ,\quad
m_g^2=\frac{-e+2\Lambda(D-3)a}{2a+c}\, ,\quad
8\pi G_{\rm eff}=\frac{1}{4 e-8a\Lambda (D-3)}\, .
\end{equation}
As we mentioned before, $\Lambda$ is the curvature scale of the background, which in the context of the present paper we write as $\Lambda=-f_{\infty}/L^2$. Now, a simple adaptation of Eqs. (2.22) and (2.23) in \cite{Aspects} yields the following general relations
\begin{eqnarray}
L^2\mathcal{L}'(f_{\infty})&=&-2eD(D-1)\, ,\\
L^4\mathcal{L}''(f_{\infty})&=&4D(D-1)\left(a+bD(D-1)+c(D-1)\right)\, .
\end{eqnarray}
As we can see from \req{kafka}, theories with Einstein-like spectrum --- \ie those for which $m_g^2=m_s^2=+\infty$ and hence \req{lineareqs} reduces to $\mathcal{E}_{ab}^L=G_{ab}^L/(8\pi G_{\rm eff})$ --- satisfy the constraints $c=-2a$, $b=a/2$. Taking these relations into account, as well as the definition of $h(f_{\infty})$ in \req{hinf}, it follows  that $G_{\rm eff}=-G/h'(f_{\infty})$ for this class of theories, as anticipated in the main text.

\section{Taub-NUT solutions in higher-order gravities}\label{ttaub}
On general grounds, the metric of Euclidean Taub-NUT/bolt solutions with base space $\mathcal{B}=\mathbb{CP}^{k}$ for generic higher-curvature gravities takes the form
\begin{equation}\label{taub}
ds^2=V(r)\left(d\tau+n A_{\mathbb{CP}^k}\right)^2+\frac{dr^2}{W(r)}+(r^2-n^2)ds^2_{\mathbb{CP}^k}\, .
\end{equation}
Here, $V(r)$ and $W(r)$ are functions to be determined by the field equations, and $n$ is the NUT charge.
The coordinate $\tau$ is periodic, and in order to remove the Dirac-Misner string \cite{Misner:1963fr} associated with the potential $A_{\mathbb{CP}^k}$, its period must be fixed to $\beta_{\tau}=2n(d+1)\pi$. All known examples of this kind of solution for higher-curvature gravities additionally satisfy $V(r)=W(r)$.
We restrict to this case in the following. The condition of being asymptotically locally AdS implies that the function $V(r)$ behaves as
\begin{equation}\label{asymp}
V(r)=f_{\infty}\frac{r^2}{L^2}+\mathcal{O}(1)\, , \quad {\text{when}}\,\, r\rightarrow\infty. 
\end{equation}
Then, we see that the boundary metric at $r\rightarrow\infty$ is conformally equivalent to the squashed sphere metric \req{squa}, with $\psi=\tau/(n(d+1))$ and with the squashing parameter $\varepsilon$ given by \req{n-epsilon}. Hence, in the holographic context, these metrics have the correct boundary geometries so as to describe the dual theories on squashed spheres. On the other hand, one has to impose regularity of the solution in the bulk. In general, there is a value of $r=r_{\rm \ssc H}$ such that $V(r_{\rm \ssc H})=0$, and we distinguish two qualitatively different cases. If $r_{\rm \ssc H}=n$ the solution is of the NUT type, whereas if $r_{\rm \ssc H}\equiv r_b >n$ it is a bolt. In both cases, absence of a conical singularity imposes the following condition on the derivative of $V$:
\begin{equation}\label{reg}
V'(r_{\rm \ssc H})=\frac{4\pi}{\beta_{\tau}}\, .
\end{equation}
This is the broad picture, but of course constructing actual solutions for a particular higher-order gravity is a difficult problem. As we mentioned, in the general case the solution is determined by two functions that   satisfy a highly non-linear system of equations, including higher-order derivatives. However, for the class of theories that we are considering here --  namely Einstein and Lovelock gravities, Einsteinian cubic gravity, Quasi-topological gravity, or in general, those of the Generalized Quasi-topological class -- the problem of finding solutions is drastically simplified \cite{NewTaub2}. As we have mentioned, for these $V(r)=W(r)$, and the higher-curvature equations of motion reduce, in each case, to a single third-order differential equation for $V(r)$. Interestingly, this allows for an integrable factor that effectively turns this into a second-order equation. Schematically we have
\begin{equation}\label{taubeq}
\frac{d}{dr}\mathcal{E}\left[V(r),V'(r),V''(r),r\right]=0\,\, \Rightarrow \,\, \mathcal{E}\left[V(r),V'(r),V''(r),r\right]=C\, ,
\end{equation}
where $C$ is an integration constant that in all cases is proportional to the total mass $M$ of the solution. In the case of Einstein and Lovelock gravities, the last equation is actually algebraic \cite{Dehghani:2005zm} and the resolution is trivial. In general, it is a second-order differential equation and two conditions are needed in order to specify a solution. It turns out the asymptotic condition \req{asymp} together with the regularity condition \req{reg} suffice to determine it. Expanding the equation \req{taubeq} near the cap $r=r_{\rm \ssc H}$, we find constraints that completely determine the allowed values of $M$ and $r_{\rm \ssc H}$. In the NUT case the radius $r_{\rm \ssc H}=n$ is already fixed and the regularity condition fixes the mass as a function of the NUT charge, $M=M(n)$. In the bolt case there can be several values of the radius $r_{\rm \ssc H}(n)$ for the same value of $n$, and for all of them we obtain as well $M(n)$. For each one of these sets of parameters $\{ M(n),\, r_{\rm \ssc H}(n)\}$, the function $V(r)$ can be constructed from $r=r_{\rm \ssc H}$ to infinity by using numerical methods, and we always find that for a given $\{ M(n),\, r_{\rm \ssc H}(n)\}$ it is unique.  A remarkable feature of these theories is that the thermodynamics of the solutions can be characterized fully analytically. Indeed, it is also possible to compute exactly the free energy by evaluating the corresponding regularized Euclidean actions. This is illustrated in appendix \ref{checks}.

\section{Explicit checks of formula \req{fee0e}}\label{checks}

We have verified that our conjectured formula \req{fee0e} correctly reproduces the free energies of all Taub-NUT solutions known in the literature, computed using the standard on-shell action approach. This includes Einstein gravity and Gauss-Bonnet in general dimensions as well as the recently constructed solutions of Einsteinian cubic gravity and Quartic Generalized Quasi-topological gravities in $d=3$ and $d=5$ respectively.

In the case of $(d+1)$-dimensional Gauss-Bonnet, the complete Euclidean action, including the generalized Gibbons-Hawking boundary term \cite{Gibbons:1976ue, Myers:1987yn, Teitelboim:1987zz} and counterterms \cite{Balasubramanian:1999re, Brihaye:2008xu} reads
\begin{align} 
I_E^{\rm GB} =& - \int \frac{d^{d+1} x \sqrt{g}}{16 \pi G} \left[\frac{d(d-1)}{L^2} + R + \frac{\lambda_{\ssc \rm GB} L^2 {\cal X}_4}{(d-2)(d-3)}  \right] - \frac{1}{8 \pi G} \int_{\partial} d^d y \sqrt{h} \left[K + \frac{2 L^2 \lambda_{\ssc \rm GB}}{(d-2)(d-3)} \left[{\cal J} - 2 {\cal G}_{ij}K^{ij} \right] \right]\, ,
\nonumber\\
&- \frac{1}{8  \pi G} \int_\partial d^d y\sqrt{h} \bigg\{- \frac{(d-1)(\fin + 2)}{3 L \fin^{1/2}} - \frac{L(3\fin - 2) \Theta[d-3]}{2 \fin^{3/2} (d-2)} {\cal R} 
\nonumber\\
&- \frac{L^3 \Theta[d-5]}{2 \fin^{5/2}(d-2)^2(d-4)}\left[(2-\fin) \left({\cal R}_{ij}{\cal R}^{ij} - \frac{d}{4(d-1)}{\cal R}^2 \right) - \frac{(d-2)(1-\fin)}{d-3} {\cal X}_4^{(h)} \right] + \cdots \bigg\} \, ,
\end{align}
where ${\cal X}_4 = R_{abcd}R^{abcd} - 4 R_{ab}R^{ab} + R^2$ is the Gauss-Bonnet density, $K_{ij}$ is the extrinsic curvature of the boundary with $K = h^{ij}K_{ij}$ its trace, ${\cal J} = h^{ij}{\cal J}_{ij}$ with
\begin{equation} 
{\cal J}_{ij} = \frac{1}{3} \left(2 K K_{ik}K^k_j + K_{kl}K^{kl}K_{ij} - 2 K_{ik}K^{kl}K_{lj} - K^2 K_{ij} \right)\, ,
\end{equation}
and ${\cal G}_{ij}$ is the Einstein tensor of the boundary metric $h_{ij}$.  We have also explicitly included the counterterms that ensure a finite on-shell action for  $d < 7$. The dots stand for additional contributions that are required in higher-dimensions.  Computing the on-shell action of Taub-NUT solutions in this theory yields 
\begin{align}
F^{\rm EGB}_{\mathbb{S}_{\varepsilon}^{d}}=\frac{(-1)^{\frac{(d-1)}{2}}\pi^{\frac{d}{2}}(1+\varepsilon)^{\frac{(d+1)}{2}}d(d-1)L^{d-1}}{16\Gamma\left[\frac{d+2}{2}\right]f_{\infty}^{\frac{(d+1)}{2}} G}\left[1-\frac{f_{\infty}(d+1)}{(d-1)(1+\varepsilon)}+\frac{(f_{\infty}-1)(d+1)}{(d-3)(1+\varepsilon)^2}\right]\, ,
\end{align}
which is in precise agreement with the result obtained using the conjectured relationship~\req{fee0e}.

Our next example is ECG plus a quartic generalized quasi-topological term in $d = 3$. The Euclidean action with generalized boundary and counterterms reads~\footnote{Note that here we use the simple method for generating generalized boundary and counterterms introduced in~\cite{HoloECG}. There it was found that for Einstein-like higher-order gravities a finite on-shell action for asymptotically AdS spaces is obtained by using the Gibbons-Hawking-York boundary term along with the counterterms for Einstein gravity all weighted by $a^*$ --- c.f. Eq.~(4.19) of that work.}
\begin{equation}
\begin{aligned}
I_E=-\int \frac{d^4x \sqrt{g}}{16\pi G} \left[\frac{6}{L^2}+R-\frac{\mu L^4}{8} \mathcal{P}-\frac{\xi L^6}{16}\mathcal{Q} \right]
-  \frac{(1 + 3 \mu \fin^2 + 2 \xi \fin^3)}{8 \pi G}\int_\partial d^3x\sqrt{h}\left[K-\frac{2\sqrt{f_{\infty}}}{L}-\frac{L}{2\sqrt{f_{\infty}}}\mathcal{R}\right]\, ,
\end{aligned}
\end{equation}
where 
\begin{align}
\mathcal{P}= &\,12 R_{a\ b}^{\ c \ d}R_{c\ d}^{\ e \ f}R_{e\ f}^{\ a \ b}+R_{ab}^{cd}R_{cd}^{ef}R_{ef}^{ab}-12R_{abcd}R^{ac}R^{bd}+8R_{a}^{b}R_{b}^{c}R_{c}^{a} \, ,
\nonumber\\
\mathcal{Q}=&-44R^{abcd }R_{ab }^{\ \ ef }R_{c\ e}^{\ g \ h}R_{d g f h}-5 R^{abcd }R_{ab }^{\ \ ef }R_{ce}^{\ \ gh}R_{d f g h}+5 R^{abcd }R_{abc }^{\ \ \ \ e}R_{f g h d}R^{f g h}_{\ \ \ \ e}+24 R^{ab }R^{cd ef }R_{c\ e a}^{\ g}R_{d g f b}\, .
\end{align}
Evaluating the on-shell action for Taub-NUT solutions we find ~\cite{NewTaub2}
\begin{equation}
F_{\mathbb{S}_\varepsilon^{3}} = -\frac{\pi L^2 (1+\varepsilon)^2}{G f_\infty^2} \left[\frac{1}{2} - \frac{f_\infty}{(1+\varepsilon)} - \frac{\mu f_\infty^3}{(1+\varepsilon)^3} - \frac{\xi f_\infty^4}{(1+\varepsilon)^4} \right] \, ,
\end{equation}
which matches precisely the results from the conjectured formula~\req{fee0e}.

As our last example, the Euclidean action with generalized boundary terms for the quartic generalized quasi-topological theories in $d=5$ is given by~\cite{NewTaub2}
\begin{align}\label{full6}
I_E =& - \int \frac{d^6 x \sqrt{g}}{16 \pi G} \left[ \frac{20}{L^2} + R + \frac{\lambda_{\rm \ssc GB} L^2 }{6} {\cal X}_4 -\frac{ \xi L^6 }{216} \mathcal{S} - \frac{ \zeta L ^6}{144} \mathcal{Z} \right] 
	\nonumber\\
&- \frac{1 - 4 \lambda_{\ssc \rm GB} \fin  + 8 (\xi + \zeta) \fin^3}{8 \pi G} \int_\partial d^5 x \sqrt{h} \left[K  - \frac{4 \sqrt{\fin}}{L} - \frac{L}{6 \sqrt{\fin}} {\cal R} - \frac{L^3}{18 \fin^{3/2}} \left( {\cal R}_{ab}{\cal R}^{ab} - \frac{5}{16} {\cal R}^2 \right)\right]
	\nonumber\\
&+ \frac{ \lambda_{\ssc \rm GB} \fin - 6 (\xi + \zeta) \fin^3}{8 \pi G} \frac{L^3}{18 \fin^{3/2}} \int_\partial d^5 x \sqrt{h} \left( 4 {\cal R}_{ab}{\cal R}^{ab} - \frac{5}{4} {\cal R}^2 + \frac{3}{2} {\cal X}^{(h)}_4 \right) \, ,	
\end{align}
where ${\cal X}_4$ is the Gauss-Bonnet density and 
\begin{align}
{\cal S} &=  992 R_{a}{}^{c} R^{ab} R_{b}{}^{d} R_{cd} + 28 R_{ab} R^{ab} R_{cd} R^{cd} - 192 R_{a}{}^{c} R^{ab} R_{bc} R - 108 R_{ab} R^{ab} R^2 
\nonumber\\
&+ 1008 R^{ab} R^{cd} R R_{acbd} + 36 R^2 R_{abcd} R^{abcd} - 2752 R_{a}{}^{c} R^{ab} R^{de} R_{bdce} + 336 R R_{a}{}^{e}{}_{c}{}^{f} R^{abcd} R_{bedf} 
\nonumber\\
&- 168 R R_{ab}{}^{ef} R^{abcd} R_{cdef} - 1920 R^{ab} R_{a}{}^{cde} R_{b}{}^{f}{}_{d}{}^{h} R_{cfeh} + 152 R_{ab} R^{ab} R_{cdef} R^{cdef} 
\nonumber\\
&+ 960 R^{ab} R_{a}{}^{cde} R_{bc}{}^{fh} R_{defh} - 1504 R^{ab} R_{a}{}^{c}{}_{b}{}^{d} R_{c}{}^{efh} R_{defh} + 352 R_{ab}{}^{ef} R^{abcd} R_{ce}{}^{hi} R_{dfhi} 
\nonumber\\
&- 2384 R_{a}{}^{e}{}_{c}{}^{f} R^{abcd} R_{b}{}^{h}{}_{e}{}^{i} R_{dhfi} + 4336 R_{ab}{}^{ef} R^{abcd} R_{c}{}^{h}{}_{e}{}^{i} R_{dhfi} - 143 R_{ab}{}^{ef} R^{abcd} R_{cd}{}^{hi} R_{efhi} 
\nonumber\\
&- 436 R_{abc}{}^{e} R^{abcd} R_{d}{}^{fhi} R_{efhi} + 2216 R_{a}{}^{e}{}_{c}{}^{f} R^{abcd} R_{b}{}^{h}{}_{d}{}^{i} R_{ehfi} - 56 R_{abcd} R^{abcd} R_{efhi} R^{efhi} \, ,
\\
{\cal Z} &=  -112 R_{a}{}^{c} R^{ab} R_{b}{}^{d} R_{cd} - 36 R_{ab} R^{ab} R_{cd} R^{cd} + 18 R_{ab} R^{ab} R^2 - 144 R^{ab} R^{cd} R R_{acbd} 
\nonumber\\
&- 9 R^2 R_{abcd} R^{abcd} + 72 R^{ab} R R_{a}{}^{cde} R_{bcde} + 576 R_{a}{}^{c} R^{ab} R^{de} R_{bdce} - 400 R^{ab} R^{cd} R_{ac}{}^{ef} R_{bdef} 
\nonumber\\
&+ 48 R R_{a}{}^{e}{}_{c}{}^{f} R^{abcd} R_{bedf} + 160 R_{a}{}^{c} R^{ab} R_{b}{}^{def} R_{cdef} - 992 R^{ab} R_{a}{}^{cde} R_{b}{}^{f}{}_{d}{}^{h} R_{cfeh} 
\nonumber\\
&+ 18 R_{ab} R^{ab} R_{cdef} R^{cdef} - 8 R^{ab} R_{a}{}^{cde} R_{bc}{}^{fh} R_{defh} + 238 R_{ab}{}^{ef} R^{abcd} R_{ce}{}^{hi} R_{dfhi} 
\nonumber\\
&- 376 R_{a}{}^{e}{}_{c}{}^{f} R^{abcd} R_{b}{}^{h}{}_{e}{}^{i} R_{dhfi} + 1792 R_{ab}{}^{ef} R^{abcd} R_{c}{}^{h}{}_{e}{}^{i} R_{dhfi} - 4 R_{ab}{}^{ef} R^{abcd} R_{cd}{}^{hi} R_{efhi} 
\nonumber\\
&- 284 R_{abc}{}^{e} R^{abcd} R_{d}{}^{fhi} R_{efhi} + 320 R_{a}{}^{e}{}_{c}{}^{f} R^{abcd} R_{b}{}^{h}{}_{d}{}^{i} R_{ehfi} \, ,
\end{align}
are two densities belonging to the quartic generalized quasi-topological family of theories~\cite{Ahmed:2017jod}.  Computing the on-shell Euclidean action for Taub-NUT solutions of this theory yields
\begin{equation}
F_{\mathbb{S}_\varepsilon^{5}} = \frac{\pi^2 L^4 (1+\varepsilon)^3}{ G f_\infty^3 } \left[\frac{2}{3} - \frac{f_\infty}{1+\varepsilon} + \frac{2 \lambda_{\ssc \rm GB} f_\infty^2}{(1+\varepsilon)^2} - \frac{2 (\xi + \zeta) f_\infty^4}{(1+\varepsilon)^4} \right] 
\end{equation}
which, again, precisely matches the result obtained using the conjectured relationship~\req{fee0e}.

Finally, let us note that on-shell Euclidean actions for Taub solutions in Einstein gravity and Gauss-Bonnet gravity have been previously computed in~\cite{Clarkson:2002uj, KhodamMohammadi:2008fh}. Our results agree with those calculations up to an overall factor of $8/9$ in $d=5$, a factor of $3/4$ in $d = 7$, and more generally by a factor of
\begin{equation}
\frac{2^k k!}{(k+1)^k}
\end{equation} 
in $d = 2k+1$ dimensions. These factors are precisely the ratio of the volume of a product of $k$ 2-spheres to the volume of $\mathbb{CP}^k$. This discrepancy was observed in~\cite{Bobev:2017asb} in the case $d = 5$. Both there and in the present work properly accounting for these factors is important for matching the general results expected from field theory considerations, \eg the proportionality factor between $F_{\mathbb{S}_\varepsilon^{5}}''(0)$ and $\ctt$. This, combined with our careful analysis of the computations in~\cite{Clarkson:2002uj, KhodamMohammadi:2008fh}, gives us confidence that the results presented here are correct.

\section{Free-field calculations}\label{ffc}
The numerical results for a free (conformally-coupled) scalar field and a free Dirac fermion used in the main text were presented in \cite{Bobev:2017asb}. We quickly summarize them here, along with some further details on the manipulations we performed to produce the curves in Fig. \ref{fig.2}. 

In each case, the corresponding partition functions are given, for a generic background metric, by
\begin{equation}
Z^{\rm s}=\int \mathcal{D}\phi \, e^{-\frac{1}{2}\int d^3 x \sqrt{g} \left[(\partial \phi)^2+\frac{R \phi^2}{8} \right]}\, ,\quad 
Z^{\rm f}=\int \mathcal{D}\psi \, e^{-\int d^3 x \sqrt{g} \left[\psi^{\dagger} \left(i \slashed{D} \right)\psi  \right]}\, ,
\end{equation}
where $\phi$ and $\psi$ stand, respectively, for a bosonic scalar field and a Dirac fermion. Also, $R$ stands for the Ricci scalar of $g_{\mu\nu}$ and $\slashed{D}$ is the Dirac operator on the corresponding background. For a given background geometry $\mathcal{M}$, the free energy of these fields can be written in a unified way as
\begin{equation}
F_{\mathcal{M}}=\frac{(-1)^f}{2^{(f-1)}}\log \det \left[\mathfrak{D}_{\mathcal{M}}/\Lambda^f \right]\, .
\end{equation}
Here,  $\Lambda$ is an energy cutoff, $\mathfrak{D}_{\mathcal{M}}$ stands for the conformal Laplace operator or the Dirac operator in each case, and $f=1,2$ for the fermion and the scalar respectively. Using a heat-kernel regulator \cite{Anninos:2012ft,Vassilevich:2003xt}, one can write the above expression as  
\begin{equation}
\log \det \left[\mathfrak{D}_{\mathcal{M}}/\Lambda^f \right]=\sum_i \int_{1/\Lambda^2}^{\infty}\frac{dt}{t}e^{-t\lambda_i^{3-f}}\, ,
\end{equation}
where $\lambda_i$ are the eigenvalues of $\mathfrak{D}_{\mathcal{M}}$. This expression can  in turn be split into two parts, containing UV and IR modes, respectively. Once the UV divergences are conveniently identified and regularized introducing appropriate counter-terms --- something that can be done numerically in a systematic way, as explained in~\cite{Bobev:2017asb} --- one is left with a finite and unambiguous answer for the free energy, in each case. 

\begin{figure}[t]
	\centering
	 	\includegraphics[scale=0.7]{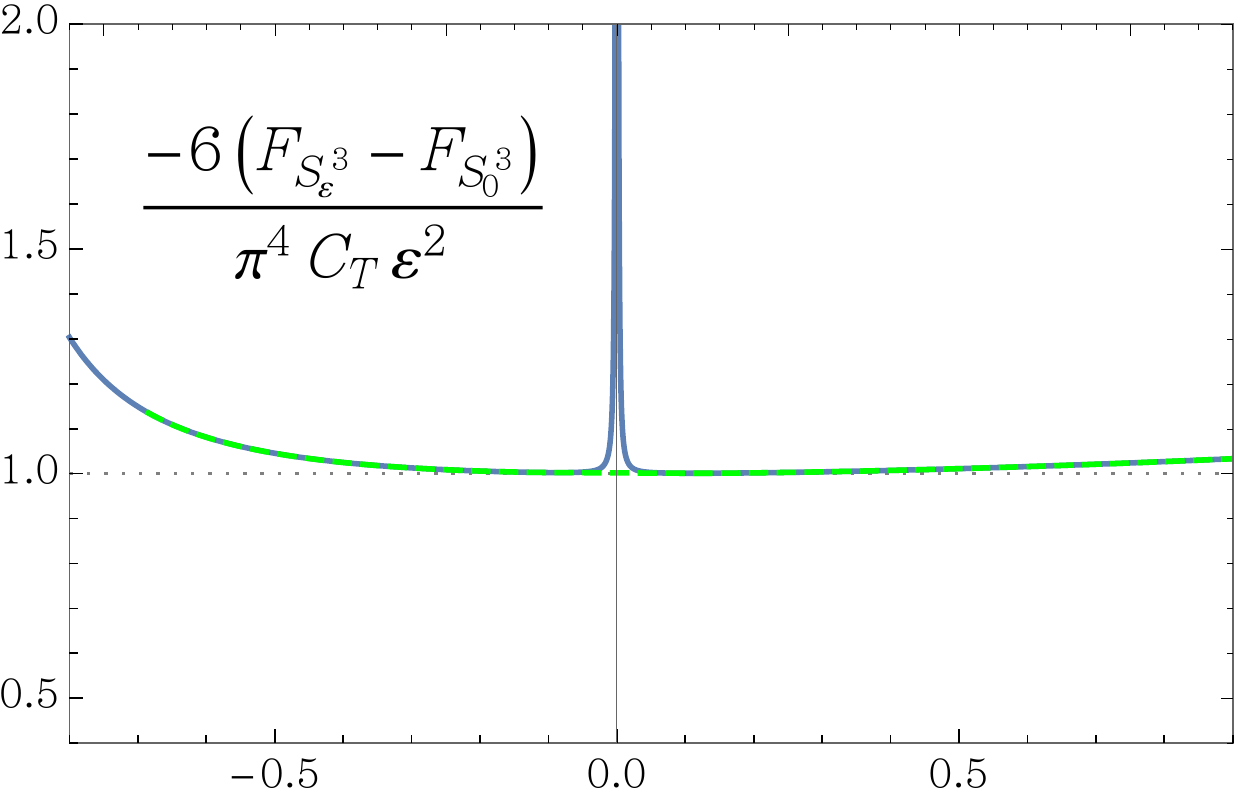}
	\caption{We plot  $-6(F_{\mathbb{S}_{\varepsilon}^{3}}-F_{\mathbb{S}_{0}^{3}})/(\pi^4\ctt \varepsilon^2)$ as a function of $\varepsilon$ using the numerical results in \cite{Bobev:2017asb}. In general, this function must cross the $\varepsilon=0$ axis at $1$, which requires a small treatment (resulting in the dashed green line) of the numerical data in a small neighborhood of $\epsilon=0$.}
	\label{fign}
\end{figure}

Plots of the numerical results obtained for $F^{\rm f}_{\mathbb{S}_{\varepsilon}^{3}}$ and $F^{\rm s}_{\mathbb{S}_{\varepsilon}^{3}}$ as functions of the squashing parameter $\varepsilon$ can be found in~\cite{Bobev:2017asb}.
Here we would like to make a technical comment about the procedure followed to obtain the curves appearing in Fig. \ref{fig.2}. Naturally, the idea is to plug the numerical results for $F^{\rm f}_{\mathbb{S}_{\varepsilon}^{3}}$ and $F^{\rm s}_{\mathbb{S}_{\varepsilon}^{3}}$ into the function $T(\varepsilon)$ and identify the value $T(0)$. In practice, this procedure requires a small treatment of the data near $\varepsilon=0$. The issue comes from the fact that $T(\varepsilon)$ involves dividing numerical expressions by $\varepsilon$, which produces divergences very close to $\varepsilon=0$. For example, we know that, for any CFT, the $\varepsilon=0$ limit of $-6(F_{\mathbb{S}_{\varepsilon}^{3}}-F_{\mathbb{S}_{0}^{3}})/(\pi^4\ctt \varepsilon^2)$ must be equal to one. However, the interpolating curves we obtain from the numerical data do not exactly account for the divergent piece in the denominator, which produces a spurious behavior very close to $\varepsilon=0$. This is illustrated in Fig. \ref{fign} in the case of the scalar. Luckily, the issue appears only within a very small neighborhood of $\varepsilon=0$, and we can safely correct these numerical values with a simple interpolation without losing any physical information about the tendency of the function in that region. Taking this into account, the functions $T(\varepsilon)$ can be safely constructed without polluting the numerical data, producing compelling evidence in favor of our general conjecture in \req{3conj}.

\bibliographystyle{apsrev4-1} 
\vspace{1cm}
\bibliography{Gravities} 

\end{document}